\documentclass[12pt,aps,prd,preprint,tightenlines,superscriptaddress,showpacs,nofootinbib]{revtex4}
\usepackage{amsmath,amssymb,amsthm,amsfonts}
\usepackage{slashed}
\usepackage{graphicx}
\usepackage{subfigure}
\usepackage{dcolumn}
\usepackage{hyperref}      
\usepackage{bm}
\usepackage{epsfig}
\usepackage{setspace}
\newcommand{\Eqref}[1]{Eq.~(\ref{#1})}

\newcommand{\secref}[1]{Sec.~\ref{sec:#1}}

\newcommand{\appref}[1]{Appendix~\ref{sec:#1}}
\newcommand{\figref}[1]{Fig.~\ref{fig:#1}}
\newcommand{\figsref}[2]{Figs.~\ref{fig:#1} and \ref{fig:#2}}
\newcommand{\tableref}[1]{Table~\ref{table:#1}}

\newcommand{\beq}{\begin{equation}}
\newcommand{\eeq}{\end{equation}}
\newcommand{\beqa}{\begin{eqnarray}}
\newcommand{\eeqa}{\end{eqnarray}}
\newcommand{\nn}{\nonumber}

\begin{document}
\preprint{NSF-KITP-11-188}
\title{Kinematic Edges with Flavor Splitting and Mixing \vspace{2pt} }

\author{Iftah Galon}
\affiliation{Physics Department, Technion---Israel Institute of 
Technology, Haifa 32000, Israel} 
\author{Yael Shadmi}
\affiliation{Physics Department, Technion---Israel Institute of 
Technology, Haifa 32000, Israel} 
\affiliation{Kavli Institute for Theoretical Physics,
 University of California,
 Santa Barbara, CA 93106}

\begin{abstract}
\vspace{10pt}
The kinematic edges of invariant mass distributions
provide an important tool for the possible measurements of 
superpartner masses in supersymmetric models with a neutralino LSP.
We examine the effect of lepton flavor dependence
on the kinematic endpoints of the di-lepton invariant mass distribution,
with the leptons being electrons and muons.
In the presence of slepton mass splitting and mixing,
each of these distributions exhibits multiple edges, which are
likely to be close. Furthermore, flavor subtraction, which is
usually employed to eliminate backgrounds, dilutes the signal.
We propose to extract the endpoints from the flavor-added 
distribution, which is insensitive to the slepton mixing.
We also discuss the extraction of the slepton flavor parameters
in such scenarios.
To demonstrate our results, we use an example with a small slepton mass
splitting of 3 GeV leading to a 6 GeV edge splitting, at both small mixing
and large mixing.
\end{abstract}

\maketitle

\section{Introduction}
Kinematic edges provide one of the main tools for extracting 
superpartner masses~\cite{Hinchliffe:1996iu, Hinchliffe:1998ys,Bachacou:1999zb, 
Lester:2005je, 
Lester:2006cf, Gjelsten:2004ki, Autermann:2009js, Barr:2007hy}. 
If supersymmetry, or other types of new physics, gives rise to events with
cascade decays ending in a final state with invisible massive particles, 
then the events cannot be fully reconstructed, but various invariant-mass 
distributions  
exhibit edges whose locations depend on the superpartner masses. 
Given sufficient measurements of these edges, the masses can in principle be 
inferred~\cite{Hinchliffe:1998ys}.

The best studied kinematic edge is the endpoint in the
invariant-mass distribution of opposite sign (OS) electrons and muons from
the decay of a heavy neutralino 
$\tilde \chi^0_2$ to a slepton $\tilde l$, followed by the subsequent slepton 
decay to the lightest neutralino $\chi^0_1$,
\beq\label{decay}
\tilde \chi^0_2 \to \tilde l^\pm l^\mp_j \to \tilde \chi^0_1  l^\mp_j  l^\pm_i .
\eeq
The endpoint in this case depends on the neutralino and slepton masses
through, 
\begin{equation}\label{ep}
m_{ll}^2 | _{endpoint} = \frac{(m_{\tilde\chi_2^0}^2 - m_{\tilde l}^2)(m_{\tilde l}^2 
- m_{\tilde\chi_1^0}^2)}{m_{\tilde l}^2}\,.
\end{equation}
Most studies of  kinematic edges have assumed universal slepton masses,
such that the selectron and smuon are degenerate with no flavor mixing.
The leptons $l_i$ and $l_j$ in~\Eqref{decay} are then either both electrons
or both muons, and each of the same-flavor distributions exhibits
a single endpoint: the $e^\pm e^\mp$ ($\mu^\pm \mu^\mp$) distribution 
is only sensitive to the selectron (smuon) mass.
Furthermore, since the selectron and smuon are degenerate,
the two endpoints coincide.
These features have been used to eliminate backgrounds from uncorrelated leptons
by considering the flavor-subtracted 
invariant mass distribution~\cite{Hinchliffe:1996iu}
\begin{equation} 
N_{e^+ e^- } +  N_{\mu^+ \mu^-} - N_{e^\pm \mu^\mp} 
\label{eq:flv_sub} \,.
\end{equation}

Scalar masses, however, need not be universal. 
Many examples of models with non-universal slepton masses are 
known (see for example~\cite{Feng:2007ke, Kribs:2007ac, Nomura:2008gg, 
Shadmi:2011hs,Gross:2011gj}). 
The collider signatures of flavor-violating models have been discussed
in~\cite{ArkaniHamed:1996au,ArkaniHamed:1997km,Agashe:1999bm,Hisano:2002iy,
Bartl:2005yy,Kitano:2008en,Kaneko:2008re,
Hisano:2008ng,Esteves:2009vg,Buras:2009sg,Feng:2009bs,
Feng:2009bd,DeSimone:2009ws,Feng:2009yq,Ito:2009xy, Fok:2010vk, Abada:2010kj,
Abada:2011mg,Dreiner:2011wm}.
At low-energies, such models generically give rise to
slepton mass splittings {\sl and} some degree of slepton flavor mixing.
The reason is that theories that predict different slepton
masses typically involve some new slepton quantum number,
which determines the slepton masses.
There is then some new slepton interaction basis in addition to the 
flavor basis, and the slepton masses are not necessarily diagonal
in the flavor basis.

In the presence of both mass-splitting and mixings, each di-lepton
invariant mass distribution, with $l=e,\mu$, exhibits two or more 
edges, associated with the different slepton states.
Since the selectron-smuon
mass splitting is likely to be small, the corresponding edges
may be quite close.
Compared to the usual scenario of universal slepton masses,
the edge structure in this case is therefore less sharp.
Furthermore, the same multiple edges appear in the flavor-subtracted 
distribution of~\Eqref{eq:flv_sub}.
While this distribution still eliminates the background, it dilutes the signal
as well, since 
the signal contributes to both 
the same-flavor and different-flavor decays.

The observation of kinematic edges in the presence of flavor mixing and 
splitting is thus more challenging.
Even if an edge structure is observed,
one would like to determine whether it is a single edge or a multiple edge,
corresponding to two or more new particles with small mass splittings. 
Finally, if multiple edges are observed, one would like to extract
the flavor parameters from them. A measurement of these parameters
may provide information both on the origin of the new physics,
such as the mediation mechanism of supersymmetry breaking,
and on the underlying theory of flavor.

In this paper, we study these questions 
using a toy model in which the lightest two
sleptons are selectron-smuon mixtures. Since we are mainly interested
in the ability to resolve a small edge splitting, 
we take the  mass splitting to be roughly 3~GeV,
leading to edges that are 6~GeV apart, and
consider both small mixing and large mixing. 
The small mixing and large mixing cases are somewhat complimentary.
In the former, it should probably be possible to observe the edges
in the $ee$ and $\mu\mu$ distributions, since each one of them is dominated
by a single edge.
Indeed, the zero mixing case was studied in~\cite{Allanach:2008ib}, 
where it was argued that the slepton mass splitting can be measured
down to
${\Delta m_{\tilde l}}/{m_{\tilde l}}\sim 10^{-4}$ 
(where ${\Delta m_{\tilde l}}$ is the slepton mass splitting) 
in a $14~\rm{TeV}$ LHC with $30\rm{fb}^{-1}$ 
integrated luminosity\footnote{This conclusion  depends however on the
values of the slepton masses relative to the neutralino 
masses~\cite{Allanach:2008ib}
(see discussion in Section~\ref{sec:general}).}.
If the existence of different edge locations in the $ee$ and 
$\mu\mu$ distributions can be established, it would signal flavor 
dependence and 
provide motivation for looking for flavor mixing in the $e\mu$ distribution.
On the other hand, for large mixing,
the edges in the same-flavor distributions would be harder to measure,
but the $e\mu$ distribution should exhibit some edge structure, which would
indicate flavor mixing, and provide motivation for looking for edge splitting.
In either case, as explained above, the precise determination of the edges
would be non-trivial, because the edges are ``divided''
between the four distributions $N_{l_i^+ l_j^-}$ with $l_i, l_j=e,\mu$.

To overcome this problem, we propose to consider 
the flavor-{\sl added} distribution
\begin{equation} 
N_{e^+ e^- } +  N_{\mu^+ \mu^-} + N_{e^+ \mu^-} + N_{e^- \mu^+} 
\label{eq:flv_add} \,.
\end{equation}
This is useful because: a.~the edge locations are identical in all 
the four distributions appearing in~\Eqref{eq:flv_add} since they only depend
on the slepton masses,
b.~the mixing, which affects each of the individual flavor distributions
drops out of the flavor added distribution, 
and c.~if a small edge splitting is the result of a small mass splitting,
the two sleptons make roughly equal contributions to the flavor added
distribution.
While the flavor added-distribution~\Eqref{eq:flv_add} does not get rid of 
the background,
it does not dilute the signal contributing to the edges,
and could therefore exhibit a clearer edge structure than each
of the separate flavor combinations.
Once the edge locations are measured from~\Eqref{eq:flv_add},
one can proceed to determine the mixing from the separate invariant 
mass-distributions $N_{l_i^+ l_j^-}$.

In order to see the effect of flavor dependence on kinematic edges,
it is useful to compare the edge structures with flavor-dependence
and without it. We therefore chose as our toy model the SU3 
benchmark point~\cite{Aad:2009wy} for which the selectron-smuon kinematic edge
was carefully studied, and deformed it slightly by introducing
a small selectron-smuon mass splitting and mixing by 
hand\footnote{The SU3 benchmark point may be ruled out already by 
ATLAS~\cite{atlas} and CMS~\cite{cms}, but we are only interested
in it as a toy example for assessing the effects of flavor dependence
on the dilepton edge.}.

Throughout our discussion, we assume that the slepton widths
are much smaller than the mass splitting, so that slepton flavor
oscillations can be neglected~\cite{ArkaniHamed:1996au}. 
The effect of such oscillation on the edge structure is
examined in~\cite{Grossman:2011nh}.

The outline of this paper is as follows. In~\secref{general}
we discuss the locations of the edges, and the relative numbers of
different flavor lepton  pairs. 
In~\secref{model} we present the di-lepton
invariant mass distributions for our toy model,
and extract the end-points from the flavor-added distribution.
We discuss the extraction of the remaining flavor parameters
in~\secref{flavor}.
The spectrum of our toy model is given in \appref{SU3_spectrum} and 
the fitting functions
we use in \appref{functions}.

\section{The di-Lepton Invariant Mass Distributions with Mass Splitting and
Mixing}
\label{sec:general}
We consider models in which two of the lightest sleptons, (typically the
superpartners of the Right-Handed leptons) are selectron-smuon combinations,
with
\beqa\label{defstates}
\tilde l_1&=& \cos\theta\, \tilde e -\sin\theta\, \tilde\mu   \nn\\
\tilde l_2&=& \sin\theta\, \tilde e +\cos\theta\, \tilde\mu\ ,
\eeqa
with masses 
\beq\label{masses}
m_{{\tilde l}_1}= m_{\tilde l}\,,\ \ \ 
m_{{\tilde l}_2}=m_{\tilde l}+\Delta m_{\tilde l} \,.
\eeq
We also assume that these slepton masses are between the two 
lightest neutralino masses, so that some of the heavier neutralinos
$\tilde\chi_2$ decay via~\Eqref{decay}.
Neutralino decays via slepton $i$ result in a di-lepton mass distribution
which ends at
\begin{equation}\label{edge}
m_{ll}^2 | _{edge,i} = \frac{(m_{\chi_2^0}^2 - m_{\tilde l_i}^2)(m_{\tilde l_i}^2 - 
m_{\chi_1^0}^2)}{m_{\tilde l_i}^2}\,.
\end{equation}
For small slepton mass splitting,
the difference between the endpoints can be approximated 
by~\cite{Allanach:2008ib}
\begin{equation}
\Delta m_{ll} = m_{ll}| _{edge,2}- m_{ll}| _{edge,1}
\sim \,\frac{m_{ll}}{m_{\tilde l}}
\frac{m_{\chi_2^0}^2m_{\chi_1^0}^2-m_{\tilde l}^4}
{(m_{\chi_2^0}^2 - m_{\tilde l}^2)(m_{\tilde l}^2 - m_{\chi_1^0}^2)}
\, \Delta m_{\tilde l}\,.
\end{equation}
\begin{figure}[ht]
\centering
\includegraphics[width=0.7\textwidth]{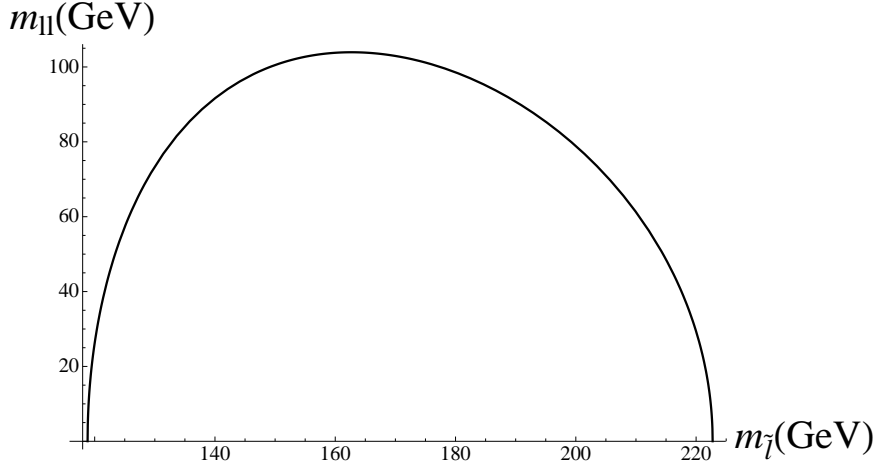}
\caption{The endpoint location, $m_{ll}|_{\rm{edge}}$, as a function of the 
slepton mass, $m_{\tilde l}$, with the neutralino masses kept fixed at 
their SU3 values of $m_{\chi_2^0} = 222~\rm{GeV}$ 
and $m_{\chi_1^0} = 118~\rm{GeV}$. }
\label{fig:mll_l}
\end{figure}
When the slepton mass $m_{\tilde l}$ 
coincides with the geometric mean of the  neutralino masses,
the edge splitting vanishes\footnote{The reason is that the maximum
of the edge~\Eqref{edge} as a function of the slepton mass occurs at
$m_{\tilde l}^2 = m_{\chi_1^0}m_{\chi_2^0}$, so around this point the sensitivity
of the edge location to the precise value of the slepton mass is 
small (see~\figref{mll_l}).}.
This has an important effect on the ability to resolve different endpoints.
For fixed neutralino masses, it would be easiest to observe
the decay~\Eqref{decay} for a slepton that is close to the geometric mean
of the two neutralino masses, since then the phase space available for 
the two emitted leptons is large, so that the leptons 
are relatively hard. 
However, for such slepton masses, the edge splitting would be
smaller than the slepton mass splitting.
On the other hand, for slepton masses far from this geometric mean,
the edge splitting can be larger than the slepton mass splittings,
but since the sleptons are closer to one of the neutralinos, 
the phase space left for either the first or the second emitted leptons is
diminished, so that this lepton is softer and therefore harder to detect.

Indeed, for the SU3 benchmark point, which was chosen partly in order
to study kinematic edges assuming selectron-smuon universality,
the slepton mass was taken to be 157~GeV, very close to 
$\sqrt{m_{\chi _2^0}m_{\chi _1^0}}\sim163$~GeV~\cite{Aad:2009wy}.
Around this mass, a small 
splitting between the sleptons could go unobserved in the edge 
structure. For example, for slepton masses varying between 
$140~\rm{GeV}$ and $185~\rm{GeV}$ the edge splitting is at 
most $5~\rm{GeV}$ as  can be seen in~\figref{mll_l}.

The numbers of events in the different di-lepton flavor contributions are
related by, 
\begin{eqnarray}
\label{eq:event_ratios}
\frac{N(e^\pm \mu^\mp)}{N(e^+e^-)} & = & 
\frac{2(1+R) \cos^2 \theta \sin^2 \theta}{\cos^4\theta + R\sin^4 \theta} 
\\ \nonumber
\frac{N(\mu^+ \mu^-)}{N(e^+e^-)} & = & \frac{R\cos^4\theta 
+ \sin^4 \theta}{\cos^4\theta + R\sin^4 \theta}\,,
\end{eqnarray}
where $R$ is the ratio of phase space factors in decays involving 
different sleptons:
\begin{equation}
R \equiv \left(\frac{m_{\chi_2^0}^2 - m_{\tilde l_2}^2}{m_{\chi_2^0}^2 
- m_{\tilde l_1}^2}\right)^2\,,
\label{eq:R}
\end{equation} 
which is close to one for near-degenerate sleptons.

\begin{figure}[ht]
\centering
\includegraphics[width=0.7\textwidth]{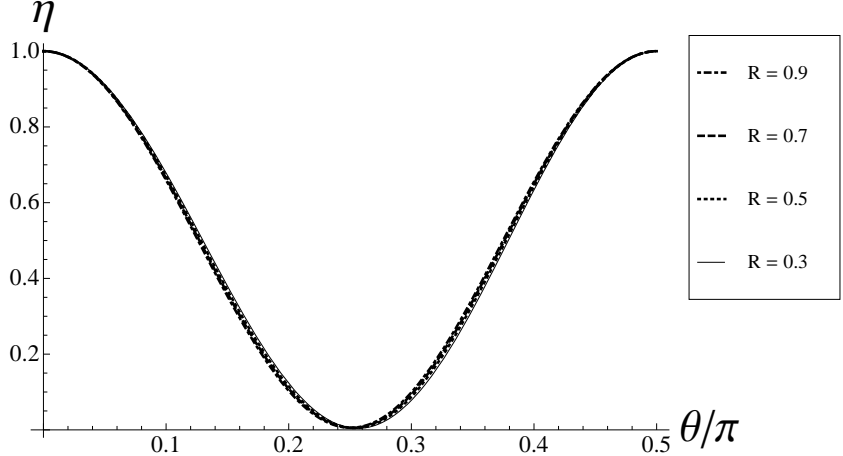}
\caption{$\eta$ (of \Eqref{eq:eta}) as a function of the mixing angle 
$\theta$, for several values of $R=0.3,0.5,0.7,0.9$, with  
$\beta=0.86$.} 
\label{fig:eta_sin}
\end{figure}

As mentioned in the Introduction, in the presence of mixing,
the flavor subtracted distribution 
dilutes the signal. 
In Figure~\ref{fig:eta_sin} we plot
the ratio of the flavor-subtracted distribution~\Eqref{eq:flv_sub}
to the total distribution,
\beqa
\eta &\equiv& \frac{N(e^+e^-)/\beta + \beta N(\mu^+ \mu^-) 
- N(e^\pm \mu^\mp)}{N(e^+e^-)/\beta 
+ \beta N(\mu^+ \mu^-) + N(e^\pm \mu^\mp)} \nn\\
&=&
1 - \frac{\beta \sin ^2 2\theta }
{\left( \beta + \sin ^2 \theta (1 - \beta) \right)^2 +
 \left( \frac{1 - \beta ^2}{1+R} \right) \cos 2\theta} 
\label{eq:eta}
\eeqa
as a function of the mixing for different values of $R$.
Here $\beta$ is the ratio of electron efficiency to muon efficiency
in the experiment. 
The weak $R$-dependence in Figure~\ref{fig:eta_sin} is a result of the 
fact that we took the ATLAS value, $\beta=0.86$ 
which is close to one. 
As expected, $\eta$ vanishes for maximal mixing, but even for 
a mixing of $\sin\theta \simeq 0.3$ $\eta$ drops to $\sim 0.6$.

\section{Toy models and results}
\label{sec:model}
\subsection{Model parameters and simulation}
As mentioned above, we use two toy models based on the SU3 benchmark point,
and modify the right-handed selectron and smuon states.
The two lightest neutralino masses are 118~GeV and 222~GeV.
The remaining masses are given in~\tableref{SU3_spectrum} of 
Appendix~\ref{sec:SU3_spectrum}.
Based on the discussion of the previous section,
we want the slepton masses to be sufficiently far from the 
geometric mean of the two neutralino masses $\sim 160$~GeV, 
so that the effect of a small slepton mass splitting is not
suppressed in the edge splitting,
but at the same time, not too close
to the neutralino masses, so that the resulting leptons are not
too soft.
We also exclude slepton masses in the 
ranges 
$135~{\rm GeV} \leq m_{\tilde l}\leq 147~{\rm GeV}$ and 
$180~{\rm GeV} \leq m_{\tilde l} \leq 196~{\rm GeV}$
in order for the edge to be separated by at least 7~GeV 
from the $Z$ resonance.

Bounds on lepton flavor violation limit the possible mass splitting
and mixing. For small mass splitting, the constrained quantity
is essentially
\beq
\label{eq:delta}
\delta_{12}^R \sim \frac{(\Delta m_{\tilde l})^2}{m_{\tilde l}^2}\sin\theta\,.
\end{equation} 
The experimental constraints on 
$\mu\to e\gamma$~\cite{Brooks:1999pu}
imply, using~\cite{Ciuchini:2007ha, Gabbiani:1996hi},
$\delta_{12}^R \le 0.09$. 

Given the considerations above, we choose the the slepton masses to be 
$m_{\tilde l_1}=131$~GeV, and $m_{\tilde l_2}=133.8$~GeV.
With these masses, the mixing is not constrained.
The resulting endpoint locations are,
\begin{eqnarray}
m_{ll}\left(m_{\tilde l_1} = 131~\rm{GeV}\right)| _{edge} = 
75.9~\rm{GeV} \nn\\
m_{ll}\left(m_{\tilde l_2} = 133.8~\rm{GeV}\right)| _{edge} = 81.9~\rm{GeV} 
\end{eqnarray}
with $\Delta m_{ll}\sim6$~GeV.
The two models we study differ only in the mixing angle. 
One has small mixing with $\sin^2\theta \simeq 0.9$,  
and the other has large mixing with $\sin^2\theta  \simeq 0.4$.

Since we are interested in a comparison of the flavor-dependent
di-lepton edges to the SU3 study, which assumed 14~TeV center-of-mass energy,
we generate $1.5 \cdot 10^5$ SUSY strong production events and 
$6\cdot 10^6$ $t\bar t$ SM events, corresponding to $10$ fb$^{-1}$ 
at a $14~\rm{TeV}$ LHC,
and use the same cuts as those used in the SU3 study~\cite{Aad:2009wy}:
\begin{enumerate}
\item Exactly two isolated opposite sign leptons (e,$\mu$) 
with $p_T > 10$~GeV and $\left|\eta\right| <2.5$.
\item At least four jets with $p_T > 50$~GeV, at least one of which has 
$p_T > 100$~GeV.
\item $\slashed E_T > 100$~GeV and $\slashed E_T > 0.2 M_{\text{effective}}$.
\item Transverse Sphericity $S_T > 0.2$.
\end{enumerate}
Based on the SU3 analysis,
other types of SM backgrounds are omitted as they become irrelevant after 
the cuts. 
We note that less than 5\% of the signal events survive the experimental 
cuts.

The spectrum for the SU3 model is calculated using
SPICE~\cite{Engelhard:2009br}, which is based on
SoftSUSY~\cite{Allanach:2001kg} and SUSYHIT~\cite{Djouadi:2006bz}.
We then modify the selectron and smuon masses, and introduce
selectron-smuon mixing by hand at low energies. 
To simulate events we use MadGraph-MadEvent (MGME)~\cite{Alwall:2007st}, 
with FeynRules~\cite{Christensen:2008py}. 
The resulting events are decayed using BRIDGE~\cite{Meade:2007js} and 
put back into MGME's Pythia-PGS 
package~\cite{MGPythia, Sjostrand:2006za, PGS} which includes
hadronization and
initial and final state radiation.
We use ROOT~\cite{Brun:1997pa} to handle the results, via MGME's 
ExROOTAnalysis package~\cite{ExROOT}.

\subsection{Resolving the Edges Using Flavor Addition}
\label{sec:flv_add}
\begin{figure}[ht]
\centering
\subfigure[~Truth distributions ($2$~GeV per bin)]{
\includegraphics[width=0.9\textwidth]{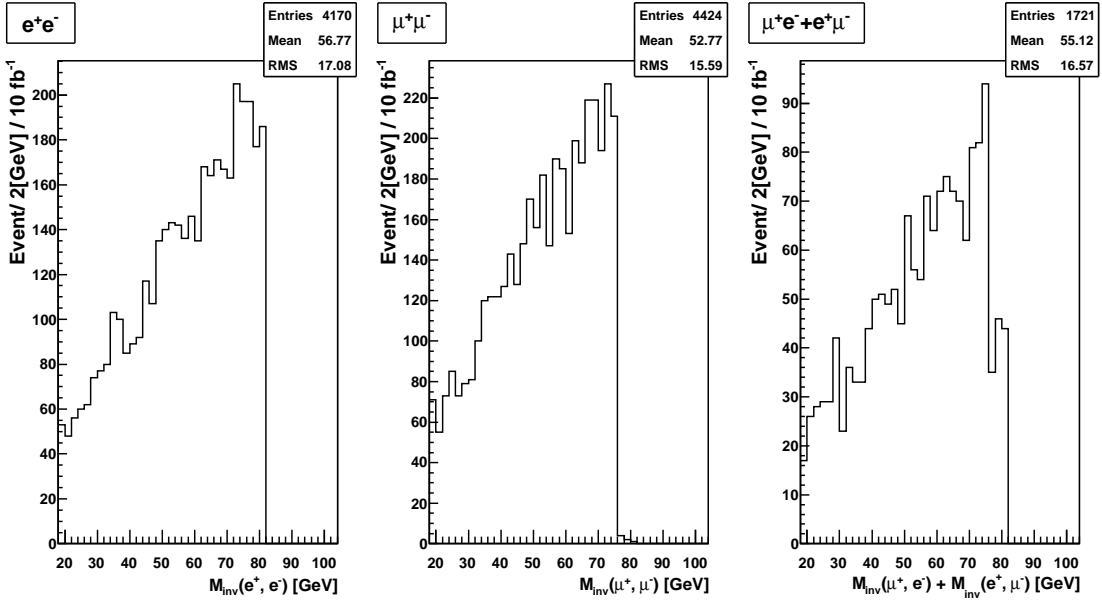}
\label{fig:dist_1truth}
}
\centering
\subfigure[~''Experimental'' distributions ($3$~GeV per bin)]{
\includegraphics[width=0.9\textwidth]{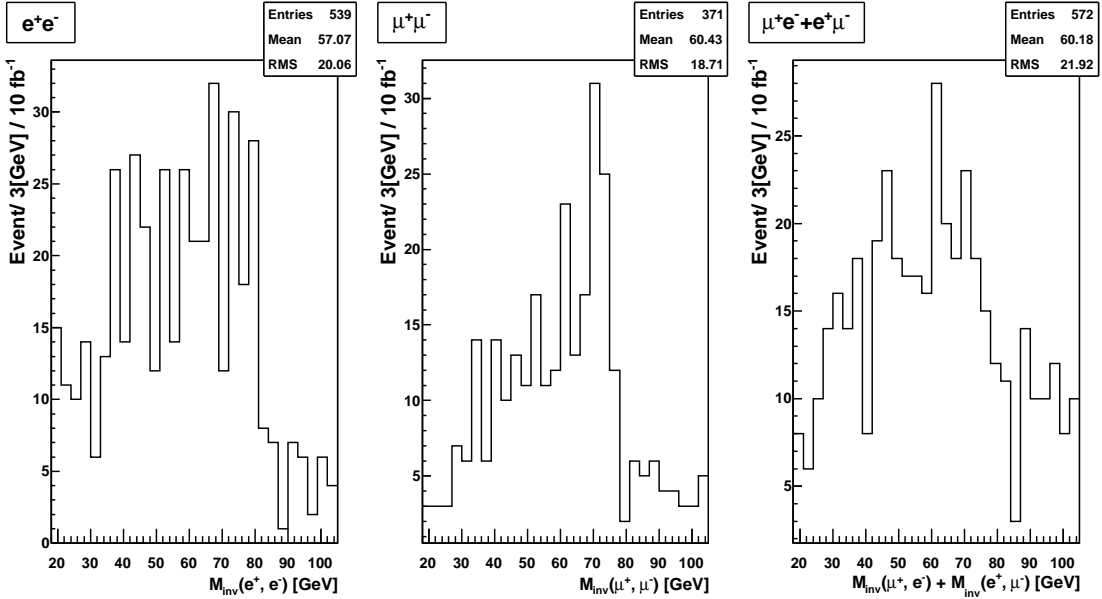}
\label{fig:dist_1exp}
}
\caption{The opposite-sign-di-lepton invariant mass distributions from 
truth---signal only before detector simulation (top), 
and ``experimental''---including background and detector simulation (bottom) 
for Model~1--small mixing.} 
\label{fig:distributions_truth_exp_1}
\end{figure}

\begin{figure}[ht]
\centering
\subfigure[~Truth distributions $2$~GeV per bin]{
\includegraphics[width=0.9\textwidth]{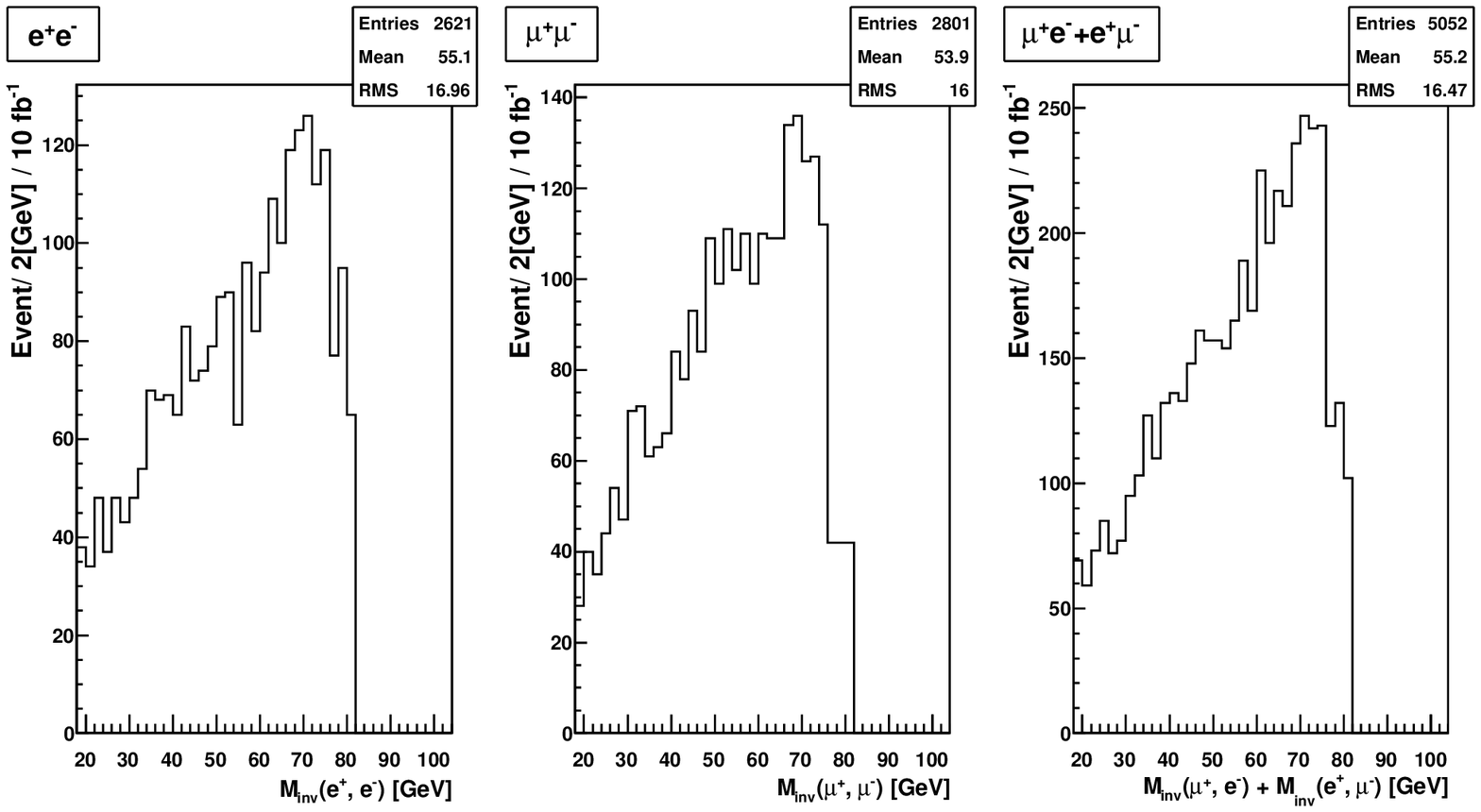}
\label{fig:dist_2truth}
}
\subfigure[~''Experimental'' distributions $3$~GeV per bin]{
\includegraphics[width=0.9\textwidth]{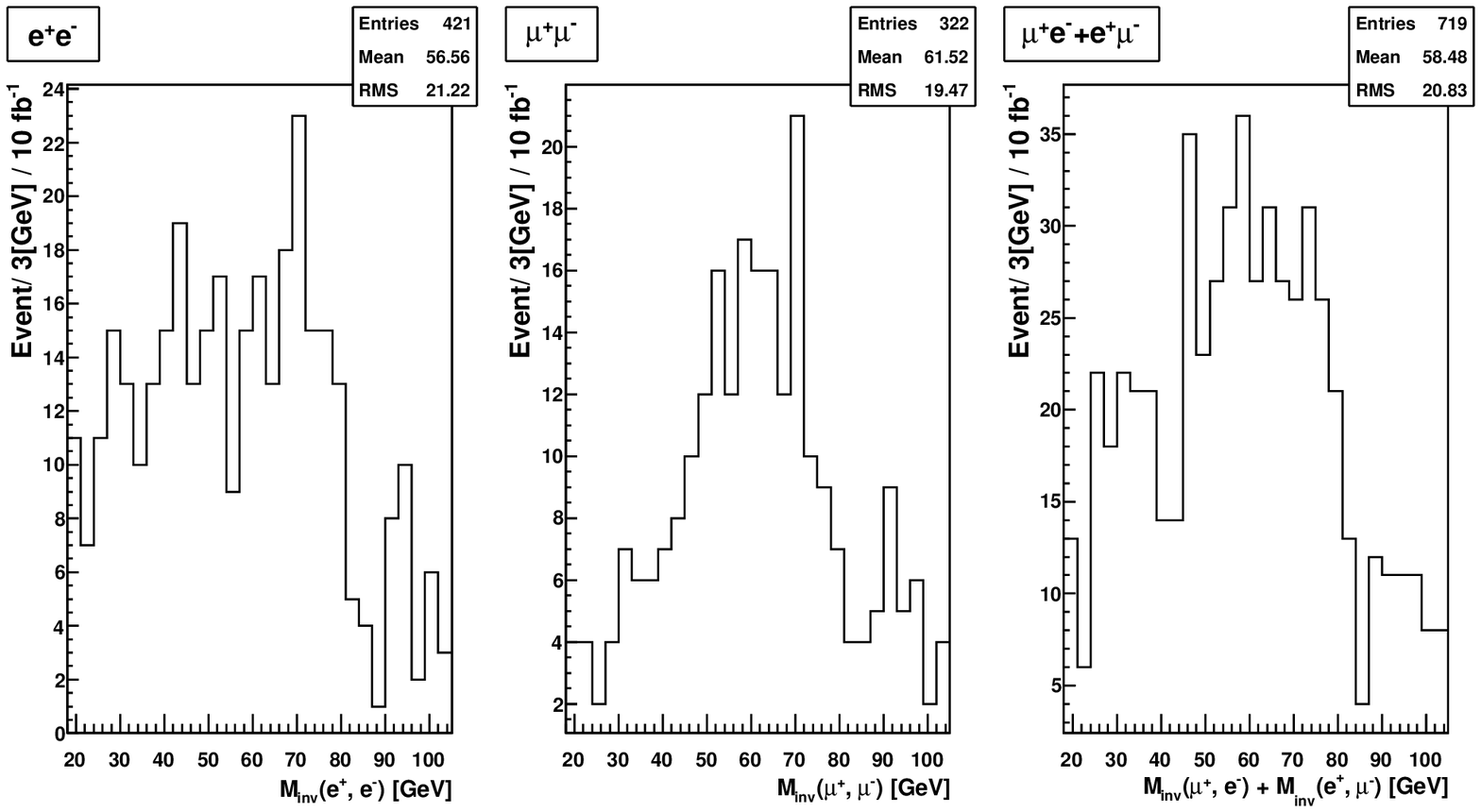}
\label{fig:dist_2exp}
}
\caption{The opposite-sign-di-lepton invariant mass distributions from 
truth---signal only before detector simulation (top), 
and ``experimental''---including background and detector simulation (bottom) 
for Model~2--large mixing.}
\label{fig:distributions_truth_exp_2}
\end{figure}

The di-lepton invariant mass distributions for the different flavor 
combinations for small mixing and for large mixing are shown in 
Figure~\ref{fig:distributions_truth_exp_1} and 
Figure~\ref{fig:distributions_truth_exp_2} respectively.
In each case, the truth distribution 
(\figref{dist_1truth}, \figref{dist_2truth})
contains the signal only,
that is, the di-leptons coming from the decay chain~\Eqref{decay}
at the generator level,
with no background from either supersymmetric events or from
top production.
The ``experimental'' distributions 
(\figref{dist_1exp}, \figref{dist_2exp})
contain both the signal and the background,
after the PGS detector simulation. 
Note that the background consists of all the possible
lepton pairs from the supersymmetric events, including leptons
from decays of charginos, $Z$ etc, as well as from SM $t\bar t$ production.
As expected, 
for small mixing, the $ee$ and $\mu\mu$ distributions
are dominated by a single slepton, and
therefore  exhibit a  single edge to a good approximation.
This edge can be easily seen in the
corresponding experimental distributions.
In contrast,
all the remaining truth distributions exhibit a double edge structure,
which translates to  a much fuzzier structure once  background
and detector effects are taken into account.

\begin{figure}[ht]
\centering
\subfigure[~Model~1--small mixing]{
\includegraphics[width=0.9\textwidth]{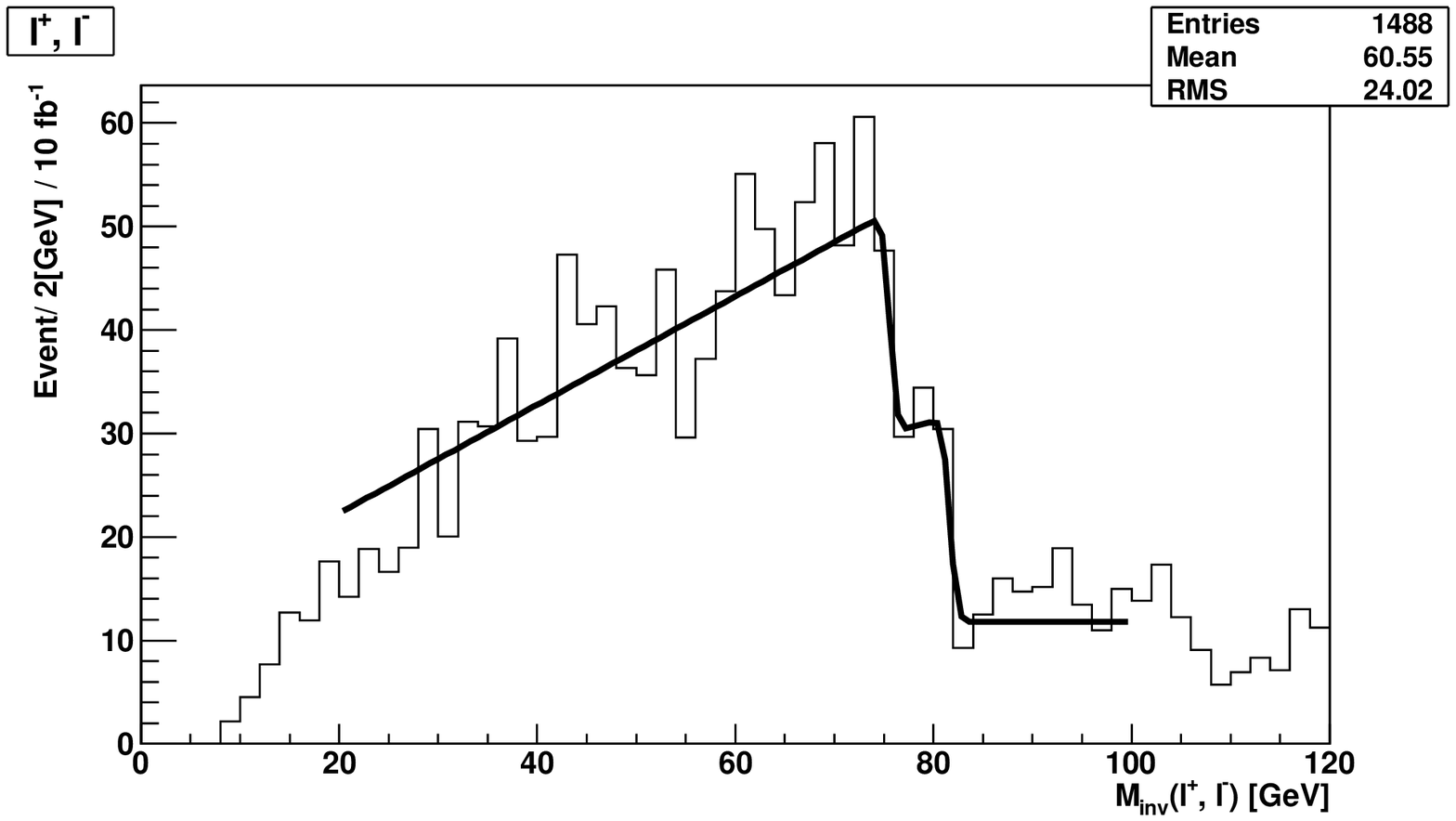}
\label{fig:comb_dist_2truth}
}
\subfigure[~Model~2--large mixing]{
\includegraphics[width=0.9\textwidth]{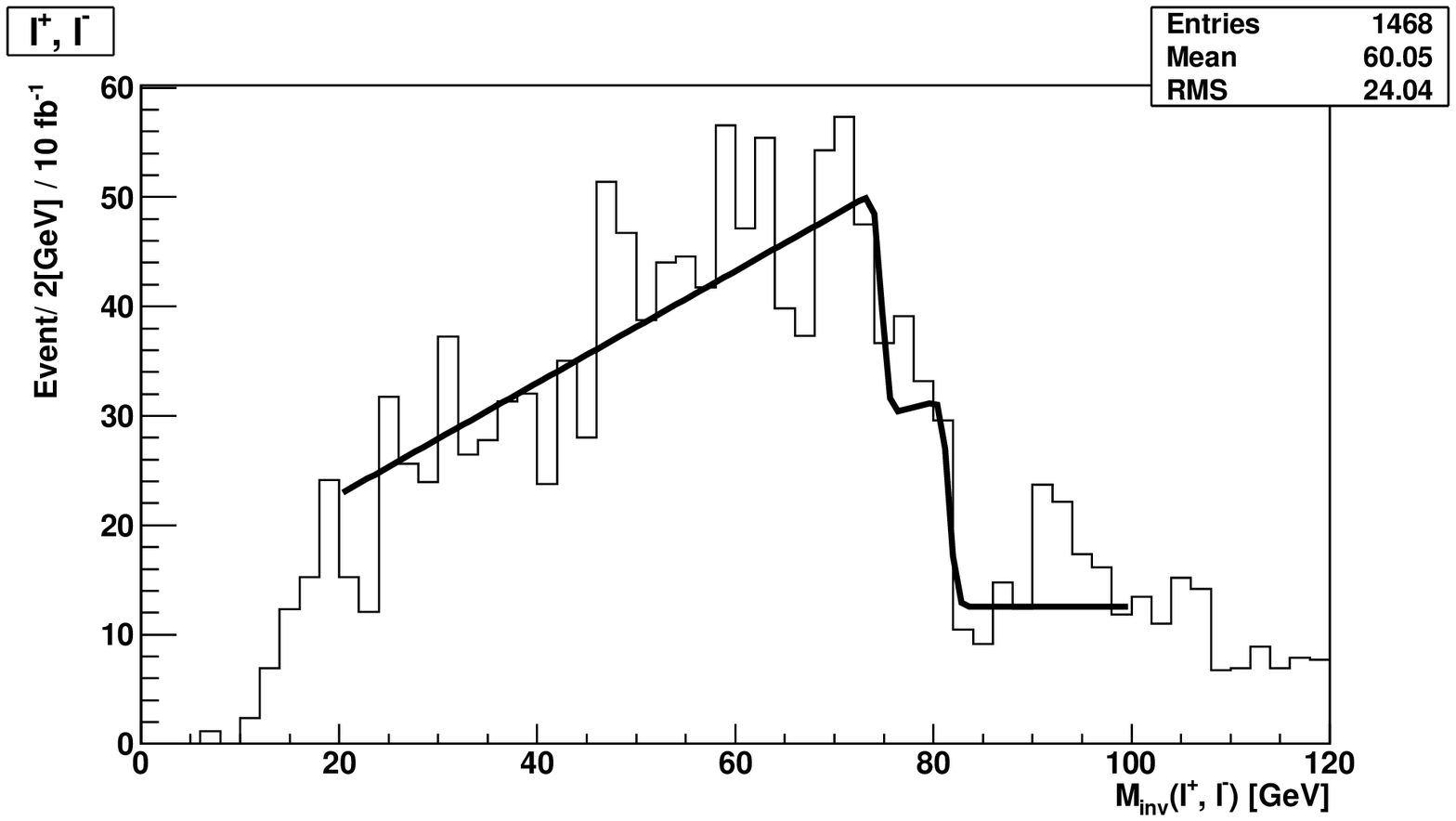}
\label{fig:comb_dist_2exp}
}
\caption{The opposite-sign-di-lepton flavor-added ``experimental'' 
invariant-mass distribution,
$N_{e^+e^-} + N_{\mu^+\mu^-}+N_{e^\pm\mu^\mp}$, for the two models at 
$2$~GeV per bin.} 
\label{fig:combined_fit}
\end{figure}

In order to obtain clearer edges we therefore exploit the fact that the
edge locations coincide in these different distributions, 
and consider the flavor-{\sl added} dilepton invariant mass distribution
$N_{l^+l^-}$ with $l=e,\mu$.
Using \Eqref{eq:event_ratios}, it is easy to see that this distribution
is independent of the mixing, with the $\tilde{l}_1$,  $\tilde{l}_2$
contributions differing by the phase space ratio $R$.
In~\figref{combined_fit}, we plot the flavor-added ``experimental'' 
invariant-mass
distributions.
As above, these contain both the signal and background,
with the background consisting of all the possible
lepton pairs from the supersymmetric events, including leptons
from decays of charginos, $Z$ etc, as well as from SM $t\bar t$ production.
Indeed, one can observe two distinct edges.
To determine the locations of these edges, we 
fit the distribution with two triangles over a constant background,
convoluted with Gaussian noise (see~\Eqref{eq:N2T} 
of \appref{functions}). 
Since the edges are clearly very close,
we can try first setting $R=1$,
for which the two slepton contributions are equal.
This is a good approximation if the small edge splitting
is the result of a small slepton mass splitting.
Expanding in $\Delta m_{\tilde l}$, it is easy to see that the deviation
of $R$ from 1 affects the distribution only at ${\cal O}(\Delta m_{\tilde l}^2)$.
However, as explained in~\secref{general},
a small edge splitting does not necessarily imply a small
mass splitting. In this case, taking $R=1$ would give a poor fit.
We will return to this case in the next section.
The four fit parameters are then the two endpoints,
the constant background, and the total number of 
events\footnote{The Gaussian noise parameter, 
$\sigma = 0.57$,
is extracted from the 
opposite-sign-same-flavor di-lepton $Z$ resonance,
which gives a reasonable approximation since it is  
close to the endpoint.}
and we find the endpoints at 75.6~GeV and 81.7~GeV
(for the small  mixing model) and at 74.8~GeV and 81.6~GeV
(for the large mixing model),
indicating that the two different edges can be resolved in this case.

The ability to resolve the edges depends of course on the binning,
which is determined in turn by the available statistics.
For coarser bins, of width larger than 6~GeV, only a single edge can
possibly be detected for our choice of model parameters.
Conversely with much higher statistics and smaller bin sizes one
can probably observe a sharper double-edge structure.
Here we wanted to focus on the trickiest scenario with the edge separation
just slightly above the statistically-significant bin size.

We note that before any fit is done, each of the distributions is scaled by 
the the proper power of $\beta$ (the ratio of electron to muon efficiencies). 
We ``measure'' the relevant efficiencies in our sample to be roughly 0.47 
for electrons,
and 0.4 for muons, with $\beta\sim 1.18$.
These low efficiencies are the result of the large number of jets
in the events, combined with the requirement of isolated leptons.
It is quite possible that
the endpoint resolution might be improved with a different set of cuts.
Thus for example, the decay chain~\eqref{decay} does not require
gluino pair production. If it originates from squark pair production,
it could be accompanied by only two jets. If only neutralinos,
charginos and sleptons can be produced at the LHC the event selection
would be completely different. With fewer jets in the final state
the efficiency for leptons would be much higher.

Equipped with the results for the endpoints from the 
flavor-added distribution, one can return to the individual flavor combinations
of \figsref{dist_1exp}{dist_2exp},
and extract additional information, starting with the mixing.
We will discuss this in \secref{flavor}.
If one has reason to believe
that the individual distributions of \figsref{dist_1exp}{dist_2exp} exhibit 
double edges,
one can of course try to simultaneously fit them with a double triangle.
This fit, however, is quite sensitive to initial conditions since it 
involves six parameters:
the total number of events, the two endpoints, $R$, the mixing 
$\sin\theta$ and the constant background.

As explained above, our anchor model was the SU3 benchmark point,
with degenerate selectron and smuon masses at $156$~GeV~\cite{Aad:2009wy}. 
The endpoint in this case was obtained using 
flavor subtraction at 
$m_{ll}|_{\text{endpoint}}=99.7 \pm 1.4|_{\rm{stat}} \pm 0.3|_{\rm{sys}}$~GeV
(the true value was $100.2$~GeV). 
The SU3 analysis considered SUSY strong production cross-sections at 
next to leading order 
$\sigma^{NLO} = 27.68(\rm{pb})$ for which 500K events were produced 
(the results were then normalized to $1\rm{fb}^{-1}$). 
Our analysis includes a more modest data sample with only leading 
order cross-sections of $\sigma^{LO} \sim 15(\rm{pb})$ 
(we give our results for $10\rm{fb}^{-1}$). 
The number of produced events and the obtained signal samples are 
however in proportion. 
Our background estimation was rather lenient and relied on the SU3 
results which indicated that the only significant contribution is 
from $t \bar t$. 
The most important ingredient of the detector simulation for our analysis
is the electron and muon efficiencies. As explained above,
these efficiencies were very low, because of the large
number of jets in the events, with a larger efficiency for 
electrons\footnote{Our overall efficiency was roughly
similar to the SU3 study~\cite{Aad:2009wy}.}.
Furthermore, the SU3 analysis used an optimized set of cuts in order to obtain
the precise endpoint locations quoted above, which we have not attempted.
Clearly then, a much better precision can be achieved for the
model we discussed here. A careful estimate of the possible
sensitivity to double edges is certainly beyond the scope
of this paper (and our ability as theorists).
Our main objective here is to examine whether edges can be detected
at all in the presence of both splitting and mixing,
and if that is the case, whether double edges can be resolved.
As we saw above, for the models we considered here, with a $\sim6$~GeV 
edge splitting, the edge structure of the separate flavor distributions 
was hard to detect,
but the flavor-added distribution indeed allowed for resolving
the endpoints.

\section{Understanding the flavor of sleptons}
\label{sec:flavor}
If an edge structure is discovered in the di-lepton invariant mass 
distribution, it would hint
at new particles that couple to electrons and muons, 
such as the slepton(s)
and neutralinos of supersymmetry.
The first ``flavor  question'' one would be faced with then is whether
there is a single ``slepton'' or multiple sleptons with similar masses.
The observation of two different edges in $N_{l_i l_j}$ would be
a clear indication of the latter. However, this may not be possible
if the two sleptons are almost degenerate, or, if their 
masses are close to 
$\sqrt{m_{\chi _2^0}m_{\chi _1^0}}\,$ so that the edge splitting is
suppressed as explained in~\secref{general}.

In this latter case,
the appearance of $R$ in~\Eqref{eq:event_ratios} provides
complementary information on the slepton masses, since it involves
different combinations of the slepton and neutralino masses.
As long as the slepton mass splitting is not too small,
the different ratios in ~\Eqref{eq:event_ratios} may provide a measurement of
$R\neq1$, and therefore establish the existence of two slepton states
with different couplings to electrons and muons.

As we argued in \secref{model}, the flavor added distribution
is very useful for measuring the endpoints, because it 
does not dilute the signal, and is independent of the mixing.
Once the endpoints are measured from this distribution,
one can turn to the individual flavor combinations
and simultaneously fit them using~\Eqref{eq:event_ratios} 
(see \appref{sim_fit}) with the values found for the
two endpoints as input.
The fit depends on four
parameters: the number of signal events, $R$, the mixing $\sin\theta$ 
and the constant background.
Performing this fit for our two toy models, we extract the
mixing reasonably well, with $sin^2\theta=0.8$ (compared to
the true value of 0.9) for the small mixing model,
and $sin^2\theta=0.5$ (compared to 0.6) for the large mixing
model. The results are collected in Table~\ref{table:results}.
\begin{table}[ht]
\subtable[Model~1--small mixing]{
\begin{tabular}{||l||c||c||c||}
\hline \hline 
Parameter              &   Truth  &   ``Measured''    & Error \\ \hline \hline
$EP1$                        &   75.86  &   75.57  & 0.76  \\ \hline \hline
$EP2$                        &   81.87  &   81.68  & 0.55  \\ \hline \hline
$R$                              &   0.95    &   1.19    &   0.13   \\ \hline \hline
$\sin^2\theta$     &   0.91    &   0.79  & 0.02  \\ \hline \hline
\end{tabular}
\label{table:results_1}
}
\subtable[Model~2 - large mixing]{
\begin{tabular}{||l||c||c||c||}
\hline \hline 
Parameter          & Truth        & ``Measured''     & Error \\ \hline \hline
$EP1$                    & 75.86        & 74.75   &  0.39  \\ \hline \hline
$EP2$                    & 81.87        & 81.61   &  0.60   \\ \hline \hline
$R$                         & 0.95           & 1.79 &  0.69  \\ \hline \hline
$\sin^2\theta$ & 0.585        &0.534    &  0.043  \\ \hline \hline
\end{tabular}
\label{table:results_2}
}
\caption{Fit results for the endpoints and flavor parameters for 
Model~1--small mixing and Model~2--large mixing. 
The errors are only the fit errors.}
\label{table:results}
\end{table}

With the value of $R$ and the two endpoints measured,
one has three different combinations
of the two slepton masses and the  two neutralino masses,
and can therefore extract three relations between these soft masses.

\section{Conclusions}
Existing tools for the  measurement of superpartner masses 
often rely on the assumption of scalar mass degeneracy.
Here we studied the effect of flavor dependence on the kinematic
edges in the di-lepton mass distribution.
It would be interesting to extend this study to distributions
involving quarks as well.

If new physics is discovered at the LHC, one would eventually like
to understand whether it exhibits any flavor dependence.
Are the new states single states, with universal couplings to different
standard model generations? Are they single states with generation-dependent
couplings? Are there three copies of new states with different or equal
masses and with different couplings to the standard model generations?
In~\cite{Feng:2007ke}, these questions were studied for the case
of supersymmetry with a meta-stable slepton, which allows for full 
reconstruction of supersymmetric events.
In this paper, we explored these questions in the more difficult
scenario of supersymmetry with a neutralino LSP, 
focusing on the 
measurements of kinematic edges in di-lepton mass
distributions.
We discussed methods for resolving double edges, and for extracting
the mixing. In particular, measurements of both the end-points
and the relative rates of the $ee$, $\mu\mu$, and $e\mu$ distributions
can yield complementary information on the slepton flavor parameters.

Finally, we note that we focused here on supersymmetric extensions
of the standard model, assuming that the supersymmetric nature of
the new particles is already established by other means.
In fact, the invariant mass distributions of dileptons from cascade
decays of new particles may provide important information
on the spins of these new particles, and thus allow for distinguishing
between various types of new physics, such as supersymmetry and extra
dimensions~\cite{Smillie:2005ar}. As is well known, 
if the intermediate particles
involved in the decay have nonzero spin, the resulting invariant
mass distribution is no longer a triangle (see, e.g.~\cite{Wang:2008sw}
and references therein). 
The smeared double edge structure that we have discussed here
could be hard to differentiate from a distribution arising
in the case of, e.g., universal extra dimensions.
Thus, the ``inverse problem''~\cite{ArkaniHamed:2005px} of distinguishing 
between different new physics
scenarios is exacerbated by flavor dependence.
It would be interesting to explore these questions further.

\section{Acknowledgments}
We thank Y.~Grossman, A.~Harel, E.~Kajomovitz, Y.~Rozen and S.~Tarem
for useful discussions.
We thank D.~Cohen for computer support.
We are grateful to J.~Alwall, B.~Fuks, M.~Reece  and especially
J.~Conway for answering many questions about MGME, FeynRules, BRIDGE and PGS.
Y.S. thanks the KITP Santa Barbara where part of this work was
completed. 
Research supported in part by the Israel Science
Foundation (ISF) under grant No.~1155/07, by the United States-Israel
Binational Science Foundation (BSF) under grant No.~2006071,
and by the National Science Foundation under 
 Grant No.~PHY05-51164.

\appendix
\section{The SU3 Spectrum}
\label{sec:SU3_spectrum}
SU3 is an mSUGRA model defined by the following boundary conditions:
\begin{equation}
m_0 = 100~\rm{GeV} \qquad  m_{1/2} 
= 300~\rm{GeV} \qquad A_0 = -300~\rm{GeV} \qquad \tan\beta = 6  \qquad \mu > 0.
\end{equation}
The resulting spectrum appears in \tableref{SU3_spectrum}.
\begin{table}[h!]
\begin{center}
\begin{tabular}{||c||c||c||c||}
\hline \hline 
Particles & Mass [GeV] & Particles & Mass [GeV] \\ \hline \hline
$\tilde \nu_1 $ & $216$ & $\tilde \chi^+_2$ & $477$ \\ \hline \hline
$\tilde \nu_2 $ & $217$ & $\tilde \chi^+_1$ & $222$ \\ \hline \hline
$\tilde \nu_3 $ & $217$ & $\tilde g$ & $718$ \\ \hline \hline
$\tilde \chi^0_4 $ & $477$ & $\tilde l_3$ & $151$ \\ \hline \hline
$\tilde \chi^0_3 $ &$462$ & $\tilde l_4$ & $231$\\ \hline \hline
$\tilde \chi^0_2 $ &$222$ & $\tilde l_5$ & $231$\\ \hline \hline
$\tilde \chi^0_1 $ &$118$ & $\tilde l_6$ & $232$\\ \hline \hline
$\tilde u^1$ & $451$ & $\tilde d^1$ & $602$\\ \hline \hline
$\tilde u^2$ & $643$ & $\tilde d^2$ &$639$ \\ \hline \hline
$\tilde u^3$ & $643$ & $\tilde d^3$ & $642$ \\ \hline \hline
$\tilde u^4$ & $664$ & $\tilde d^4$ & $642$\\ \hline \hline
$\tilde u^5$ & $664$ & $\tilde d^5$ & $668$\\ \hline \hline
$\tilde u^6$ & $664$ & $\tilde d^6$ & $668$\\ \hline \hline
$h^0$ &$110$ & $H^0$ & $513$\\ \hline \hline
$A^0$ &  $512$& $H^+$ &  $518$\\ \hline \hline
\end{tabular}
\end{center}
\caption{Spectrum of the 
SU3 Model, calculated using SPICE.}
\label{table:SU3_spectrum}
\end{table}

\section{Functions Describing Triangular Distributions}
\label{sec:functions}

\subsection{A Single Triangle Function}
For the triangle fit we use
\begin{equation}
T_1 (x, \left[ E, S\right]) =
\begin{cases}
   {0}  & {x < 0 } \\
   {2\left(\frac{S}{E^2}\right)x }  & {0 \le x \le E} \\
   {0}  & {x > E} \\
 \end{cases}
\end{equation}
Here $E$ is the endpoint and $S$ is the area of the triangle proportional 
to the total number of events in the distribution.
\subsection{A Double Triangle Function}
We describe the sum of two triangles based at zero with two different 
endpoints and slopes by:
\begin{equation}
T_2 (x, \left[ E_1, S_1, E_2, Ratio\right]) 
= T_1 (x, \left[ E_1, S_1\right]) + T_1 (x, \left[ E_2, S_2\right])
\end{equation}
where
\begin{equation}
S_2 = S_1\times Ratio
\end{equation}
and $Ratio$ is the ratio of the triangle areas.

\subsection{A Double Triangle Convoluted With A Gaussian Function}
To account for the noise in the measurement of particle momenta one 
must convolute the distributions with a Gaussian. 
The smearing parameter for our detector is measured from 
$Z\rightarrow l^+l^-$ 
and is $\sigma = 0.568$. We use:
\begin{eqnarray}
N_2T (x, \left[ \sigma, E_1, S_1, E_2, Ratio\right]) &=& 
\frac{1}{\sqrt{2\pi \sigma^2}} \int_{-\infty}^\infty dx' 
e^{-\frac{(x-x')^2}{2\sigma^2}} T_2(x', \left[E_1, S_1, E_2, Ratio\right])
\label{eq:N2T}
\end{eqnarray}
 
\subsection{Simultaneous Fit Function}
\label{sec:sim_fit}
Our ``measured'' data set consists of the 3 di-lepton invariant mass 
distributions $e^+e^-, \mu^+\mu^-, e^\pm \mu^\mp$. 
To fit them simultaneously we use:
\begin{equation}
\text{Simultaneous} (x, \left[\sigma, E_1, E_2, \sin^2\theta, R, S_{ee}, B_{ee}, B_{\mu\mu} \right]) = 
\begin{cases}
f_{ee} & \text{for ee histogram} \\
f_{\mu\mu} & \text{for mumu histogram} \\
f_{e\mu} & \text{for emu histogram}
 \end{cases}
\end{equation}
here $\sigma$ is the smearing parameter, 
$E_1$, $E_2$ are the two endpoints, 
$S_{ee}$ is the combined area of the two triangles in the $ee$ distribution, 
$B_{ll}$ is the constant background in the $ll$ distribution.
In addition,
\begin{eqnarray}
f_{ee} &=& N_2T(x, \left[
{\sigma, E_1, S_{ee}\frac{\cos^4\theta}{\cos^4\theta + R\sin^4\theta} , 
E_2, R \frac{\sin^4\theta}{\cos^4\theta} }\right]) + B_{ee} \nonumber \\
f_{\mu\mu}  &=& N_2T(x, \left[
{\sigma, E_1, S_{\mu\mu}\frac{\sin^4\theta}{\sin^4\theta + R\cos^4\theta}, 
E_2, R \frac{\cos^4\theta}{\sin^4\theta} } \right]) + B_{\mu\mu} \nonumber \\
f_{e\mu}   &=& N_2T(x, \left[
{\sigma, E_1,  S_{e\mu}\frac{1}{1+R}, E_2, R} \right]) +  B_{ee} +  B_{\mu\mu} 
\nonumber \\
\end{eqnarray}
where we have defined:
\begin{equation}
 S_{\mu\mu} =  S_{ee} \times \frac{\sin^4\theta 
+ R\cos^4\theta}{\cos^4\theta + R\sin^4\theta}
\end{equation}\begin{equation}
S_{e\mu} =  S_{ee} \times 2\frac{(1+ 
R)\sin^2\theta \cos^2\theta}{\cos^4\theta + R\sin^4\theta}
\end{equation}
so that $S_{\mu\mu}$ ($S_{e\mu}$) is the total area of the two 
triangles in the $\mu\mu$ ($e\mu$) distribution.

\bibliography{bibaFlavorEndPoints}

\providecommand{\href}[2]{#2}\begingroup\raggedright\endgroup

\end{document}